\documentclass{eas}
\usepackage{graphicx}
\begin{document}

\title{Infrared spectroscopic diagnostics for Active Galactic Nuclei} 
\author{Luigi Spinoglio}\address{Istituto di Fisica dello Spazio Interplanetario, INAF, Via Fosso del Cavaliere 100,0133 Roma, Italy}
%
%
\begin{abstract}
Infrared spectroscopy in the mid- and far-infrared provides powerful diagnostics for studying the emission regions in active galaxies. The large variety of ionic fine structure lines can probe gas conditions in a variety of physical conditions, from highly ionized gas excited by photons originated by black hole accretion to gas photoionized by young stellar systems. The critical density and the ionization potential of these transitions allow to fully cover the density-ionization parameter space. Some examples of line ratios diagrams using both mid-infrared and far-infrared ionic fine structure lines are presented. The upcoming space observatory \textit{Herschel} will be able to observe the far-infrared spectra of large samples of local active galaxies. Based on the observed near-to-far infrared emission line spectrum of the template galaxy NGC1068, are presented the predictions for the line fluxes expected for galaxies at high redshift. To observe spectroscopically large samples of distant galaxies, we will have to wait fot the future space missions, like SPICA and, ultimately, FIRI.
\end{abstract}
\maketitle

\section{Introduction}

The spectroscopic observations of the \textit{Infrared Space Observatory}
(ISO) (Kessler et al. \cite{kes96}) opened a new window for the study of the physical and
chemical properties of IR-bright, ultraluminous infrared galaxies
(ULIRG) and active galactic nuclei (AGN). Most of the pioneering work on the mid-to-far infrared spectra of active and starburst galaxies is derived from ISO spectroscopy (Sturm {\em et al.\/} \cite{stu02}; Spinoglio {\em et al.\/} \cite{spi05}; Verma {\em et al.\/} \cite{ver03}; Verma {\em et al.\/} \cite{ver05}).

\begin{figure}
\includegraphics[width=11cm]{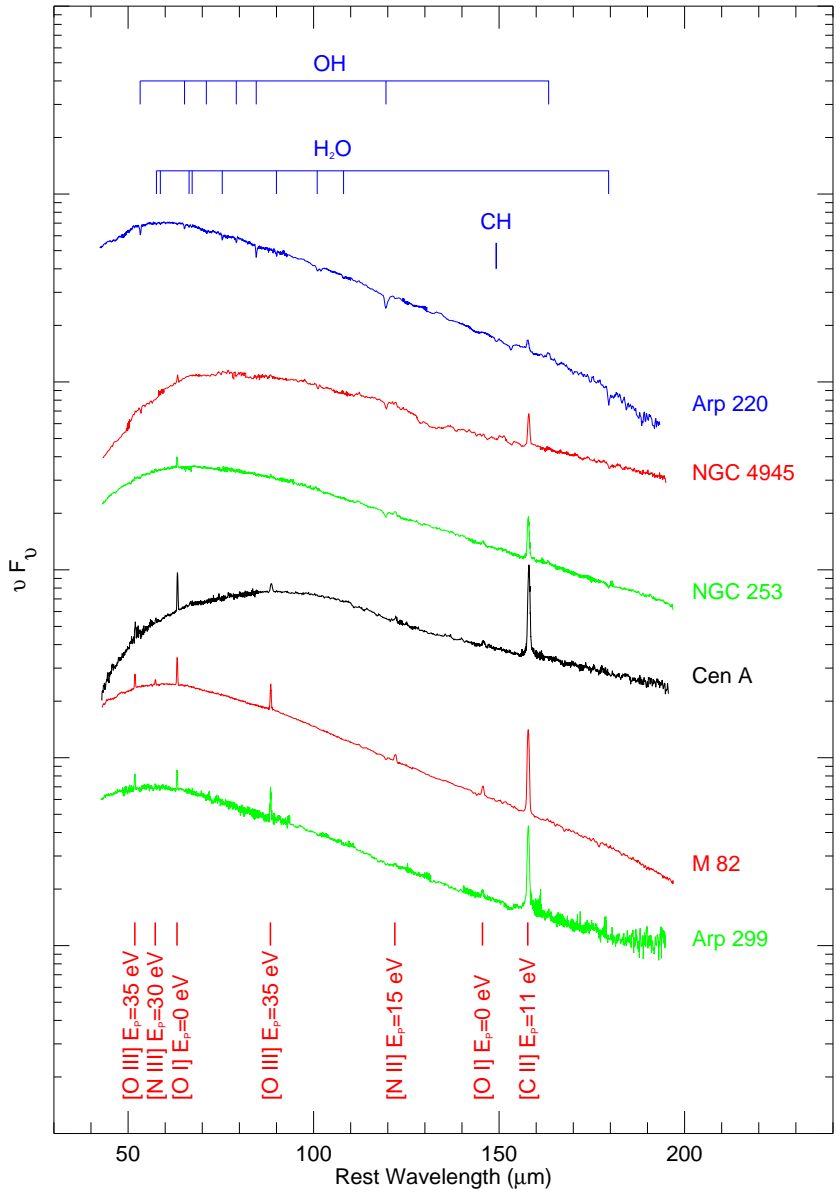}
\caption{Far infrared spectra observed with the LWS onboard of ISO of a sample of ultraluminous infrared galaxies (Fischer {\em et al.\/} \cite{Fis99})}
\end{figure}

The mid-IR spectral range includes most of the fine-structure lines excited by the hard radiation produced
by black hole accretion as well as those mainly excited by stellar
ionization (Spinoglio \& Malkan \cite{sm92}) and thus represents an essential tool
to distinguish between the two processes, especially in obscured
nuclei suffering severe dust extinction. With the advent of the \textit{Spitzer Space Telescope}, 
with its powerful mid infrared instrument, the IRS (Houck {\em et al.\/} \cite{hou04}), systematic spectroscopic 
studies of samples of galaxies have stated to appear (Dale {\em et al.\/} \cite{dal06}; Higdon {\em et al.\/} \cite{hig06}; Brandl {\em et al.\/} \cite{bra06}; Armus {\em et al.\/} \cite{arm07}; Farrah {\em et al.\/} \cite{far07}; Buchanan {\em et al.\/} \cite{buc06}; Tommasin {\em et al.\/} \cite{tom07}).

In a complementary way, the far-IR range contains a large variety of molecular (OH,
H$_{2}$O, high-J CO) and low excitation ionic/atomic transitions,
in emission or in absorption, that can reveal the geometry and
morphology of the circumnuclear and nuclear regions in galaxies.
In particular far-infrared molecular lines could trace the expected conditions of
X-UV illuminated dusty tori predicted from the unified models (Antonucci \cite{ant93} and
whose presence in type 2 active galaxies is
foreseen to reconcile the type1/type2 dichotomy.
 
 \begin{figure}
\includegraphics[width=12cm]{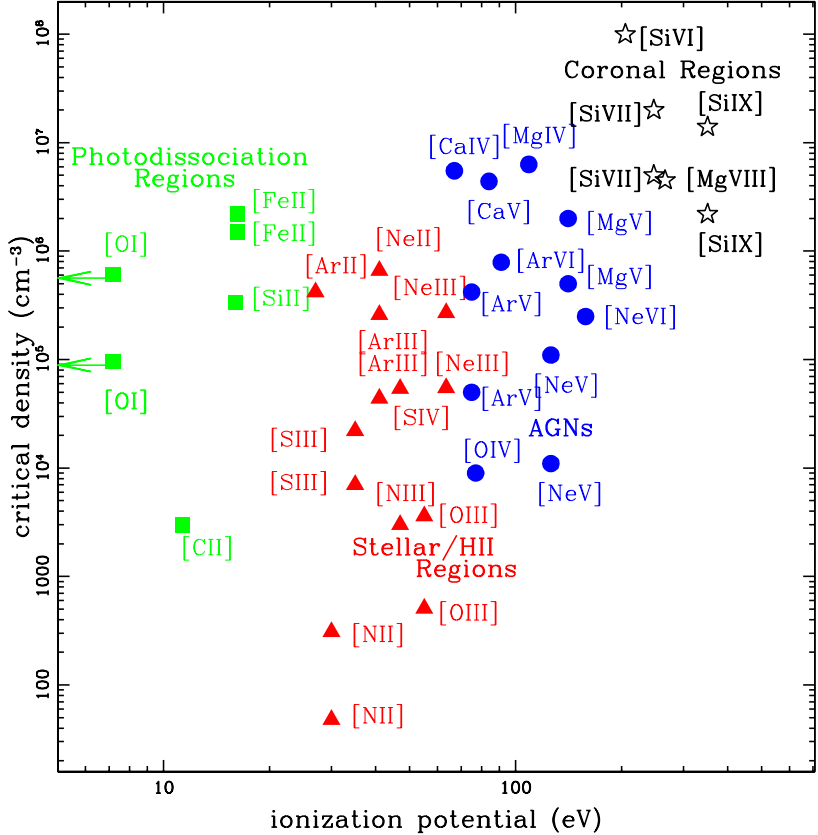}
\caption{Fine-structure lines in the 3-200 $\mu$m range. 
The lines are plotted  as a function of their
ionization potential and critical density. Different symbols are
used for lines from photodissociation regions (squares),
stellar/HII region lines (triangles), AGN lines (circles) and
coronal lines (stars). The two [OI] lines have been plotted - for
graphical reasons - at a ionization potential higher than their
effective value.}\end{figure}

The far-IR spectra of local IR-bright and ULIRG galaxies, as measured by
ISO-LWS (Fischer {\em et al.\/} \cite{Fis99}), showed an unexpected sequence of features,
as can be seen in Fig. 1, from strong [OIII]52, 88 $\mu$m and
[NIII]57 $\mu$m line emission to detection of only faint
[CII]157$\mu$m line emission and [OI]63 $\mu$m in absorption. The
[CII]157 $\mu$m line in 15 ULIRGs (L$_{IR}\geq10^{12}L_{\odot}$)
revealed an order of magnitude deficit compared to normal and
starburst galaxies relative to the FIR continuum. Non-PDR
components, such as dust-bounded photoionization regions,
generating much of the FIR continuum but not contributing
significant [CII] emission, can explain the [CII] deficiency. Such
environments may also explain the suppression of FIR
fine-structure emission from ionized gas and PAHs, and the warmer
FIR colors of ULIRGs. (Luhman {\em et al.\/} \cite{Lum03}).

LWS observations of Arp 220 show absorption in molecular lines of OH, H$_2$O, CH,
NH, and NH$_3$, as well as in the [OI]63$\mu$m line and faint
emission in the [CII]158$\mu$m line. The molecular absorption in
the nuclear region is characterized by high
excitation due to high infrared radiation density (Gonz\'alez-Alfonso {\em et al.\/} \cite{go04}).
Notably, the LWS spectrum of the prototype Seyfert 2 galaxy NGC
1068, beside the expected ionic fine structure emission lines,
shows the 79, 119 and 163$\mu$m OH rotational lines in emission,
not in absorption as in every other galaxy yet observed.
Modeling the three FIR lines of OH suggests the gas lies in small
(0.1pc) and dense clouds ($\sim 10^4 {\rm cm^{-3}}$) in
the nuclear region (potentially a signature of the torus) with a minor contribution from
the circumnuclear starburst ring at 3kpc (Spinoglio {\em et al.\/} \cite{spi05}).

\section{Fine-structure emission lines}

\begin{figure}
\includegraphics[width=12cm]{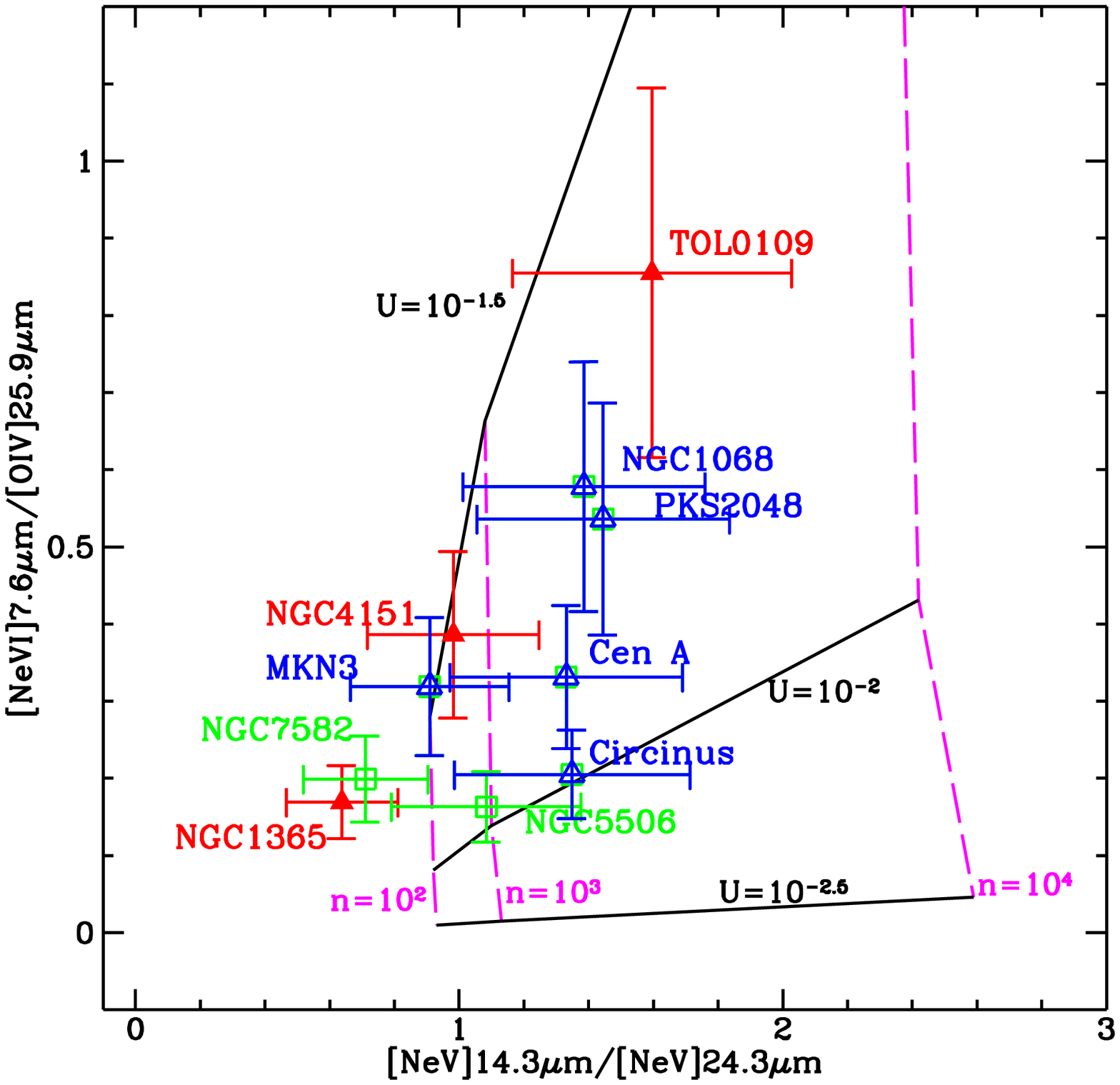}
\caption{[NeVI]7.6$\mu$m/[OIV]26$\mu$m ratio as a function of the 
[NeV]14.3$\mu$m/24.3$\mu$m ratio. The line rations of Seyfert 1's (NGC1365, NGC4151, Tol0109), Seyfert 2's (MKN3, CenA, Circinus, NGC1068, PKS2048) and NLXR galaxies (NGC5503, NGC7582) are presented, from ISO SWS observations (Spinoglio {\em et al.\/} \cite{spi00}; Sturm {\em et al.\/} \cite{stu02}). }

\end{figure}

Mid-IR and far-IR spectroscopy of fine-structure emission lines
are powerful tools to understand the physical conditions in
galaxies from the local universe to distant cosmological objects.
Fig. 2 shows the critical density (i.e. the density for which the
rates of collisional and radiative de-excitation are equal) of
each line as a function of the ionization potential of its ionic
species. This diagram shows how these lines can measure
two fundamental physical quantities (density and ionization) of
the gas. Lines from different astrophysical emission regions in
galaxies are shown in the figure with different symbols. The ratio
of two lines with similar critical density, but different
ionization potential,  gives a good estimate of the ionization,
while the ratio of lines with similar ionization potential, but
with diffrerent critical density, can measure the density of the
gas in the region (see, e.g., Spinoglio \& Malkan \cite{sm92}).
From Fig. 2 it is clear that infrared spectroscopy has a thorough
diagnostic power for gas with densities from about 10$^2$ cm$^{-3}$ 
to as high as 10$^8$ cm$^{-3}$ and ionization potentials up to 350 eV, 
using, to trace these extreme conditions, the so called coronal lines. 
Moreover, increasing the wavelength 
of the transition used, the spectroscopic diagnostics become more and 
more insensitive to dust extinction, and can therefore probe regions 
highly obscured at optical or even near-to-mid infrared wavelengths.

\begin{figure}
\includegraphics[width=12cm]{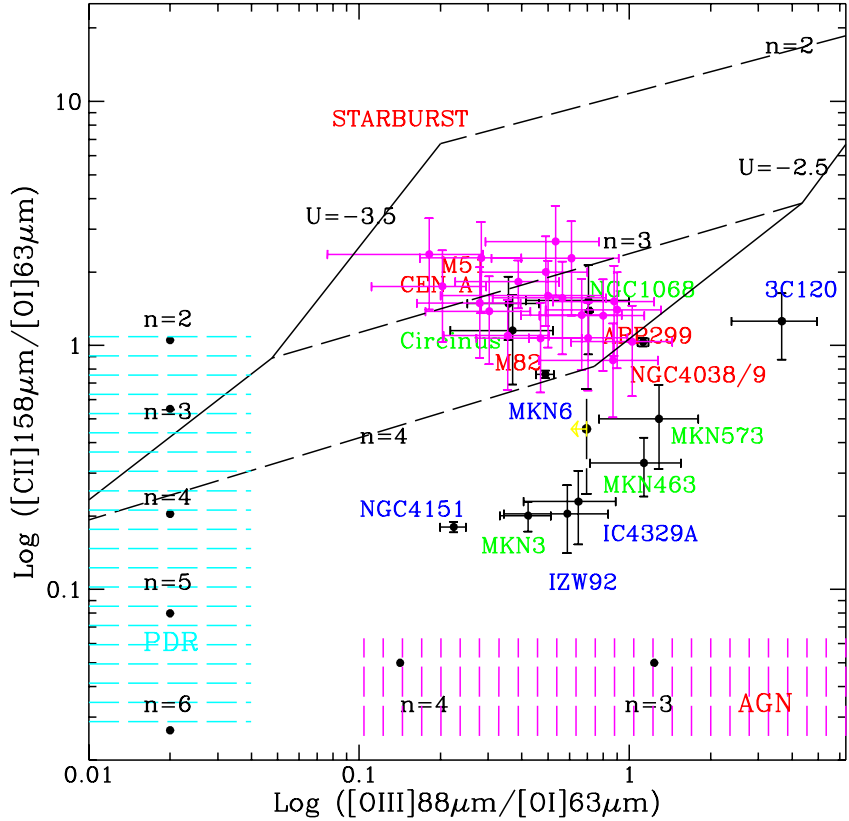}
\caption{[CII]157$\mu$m/[OI]63$\mu$m ratio as a function of the 
[OIII]88$\mu$m/[OI]63$\mu$m ratio. The grid represents starburst photoionization models computed using the CLOUDY code. At the right bottom are shown the gas density values as derived from AGN photoionization models with log U = -2.5, while at the left are given the densities derived from photodissociation region models  (Spinoglio {\em et al.\/} \cite{spi03}).}

\end{figure}

To give an example of the diagnostic power of infrared spectroscopy, we show in Fig. 3 
a diagram showing lines excited only in the highly energetic environments of AGN and not from stellar ionization. In this diagram the [NeVI]7.6$\mu$m/[OIV]26$\mu$m ratio is shown as a function of the 
[NeV]14.3$\mu$m/24.3$\mu$m ratio (Spinoglio {\em et al.\/} \cite{spi00}). The photoionization models using the CLOUDY code (Ferland \cite{fer00}) have been computed and shown as a grid in the diagram. As expected, the former ratio is sensitive to  ionization, 
while the latter is sensitive to density. In the figure are also shown measurements on a small sample of Seyfert galaxies for which we can determine the ionization potential (basically the ratio of ionizing photons over the number of hydrogen atoms) and the gas density. We note that the observed galaxies have average densities ranging from  less than 10$^2$ cm$^{-3}$ to less than 10$^4$ cm$^{-3}$ and ionization potential of 
10$^{-2.0}$ $<$ log~U $<$ 10$^{-1.5}$. This is in agreement, for the lower density objects,  with conditions of "coronal emission regions" in AGNs  (Spinoglio \& Malkan \cite{sm92}).

Another example, presented in Fig. 4,  is given by the [CII]157$\mu$m/[OI]63$\mu$m ratio as a function of the 
[OIII]88$\mu$m/[OI]63$\mu$m ratio. These low ionization lines are copiously produced in the ISM of galaxies (in photodissociation regions) and the [OIII] line is excited also in HII regions. However we can see from this diagram that while normal galaxies are clustering in a central region that can easily be explained by starburst models, most of the Seyfert galaxies are far from this locus and their rations cannot be reproduced by starburst models. They have much stronger emission of [OI]63$\mu$m than it would be expected from stellar emission only. To obtain better statistics on this diagram, we will have to wait for the \textit{Herschel} satellite, that will have the sensitivity to collect far-infrared spectroscopic observations of large samples of local galaxies.

\section{From the local to the distant Universe} 

\begin{figure}
\includegraphics[width=12cm]{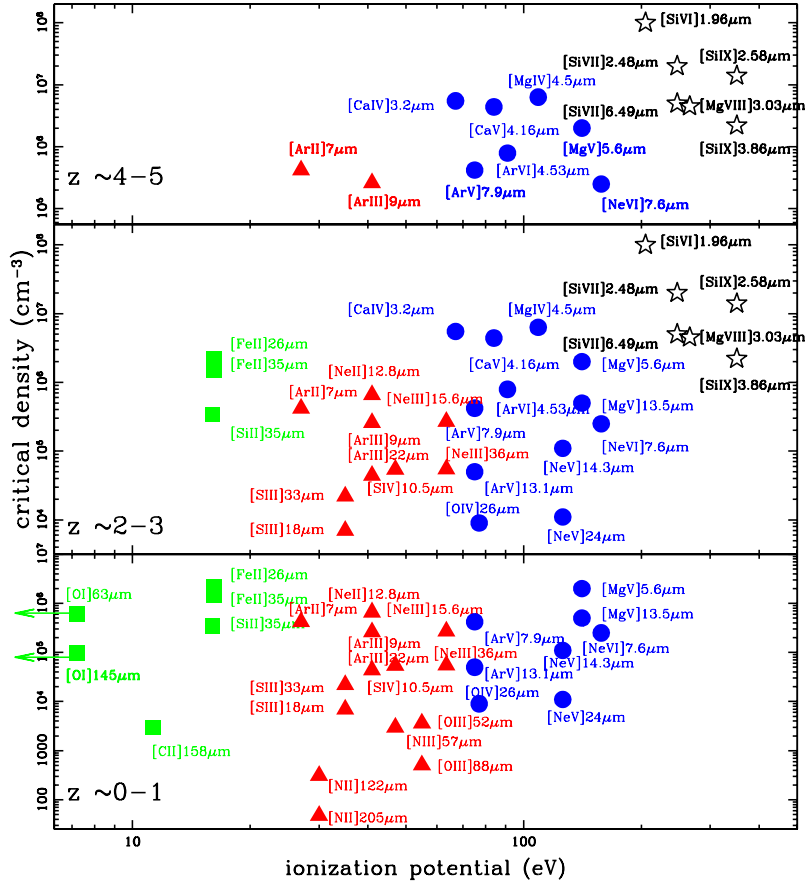}
\caption{Same as fig.1, but for differerent redshift intervals.
 Different symbols are used for lines from photodissociation regions (squares),
stellar/HII region lines (triangles), AGN lines (circles) and
coronal lines (stars). The two [OI] lines have been plotted - for
graphical reasons - at a ionization potential higher than their
effective value.}\end{figure}

Mid-IR and far-IR spectroscopy of fine-structure emission lines
can be used not only in the local universe but also to measure 
the excitation conditions in distant cosmological objects.
Fig. 5 shows again, as in Fig.2,  the critical density of the lines as a function 
of their ionization potential, not only for the local universe, but 
in three different redshift ranges, one for each frame,
for which the rest-frame wavelength of the line is shifted in the
far-infrared range. 
It appears from the figure that although the photodissociation (PDR) 
regime can be probed only in the relatively local universe, because of the long 
wavelengths of the lines tracing this regime, however, the
stellar emission (e.g. in starburst galaxies) can be probed up to
high z, using many lines in the rest-frame spectral range of 3
$\leq$ $\lambda (\mu m)$ $\leq$ 30. Moreover, the ionization from AGN can be
probed from the local universe up to redshifs of z$\leq$5 and the extremely high 
excitation coronal emission regions are probed by near-IR lines shifted into the
far-IR at a redshift of z $\sim$ 5.

Once we have proved that we have in the mid-to-far infrared the adequate diagnostic lines, we need
still to understand if these lines could be detected by the future space facilities under development or study 
in the future years. Do do so, we will use the observed infrared spectra in local galaxies and let them evolve backwards at earlier cosmological times.

\begin{figure}
\includegraphics[width=12cm]{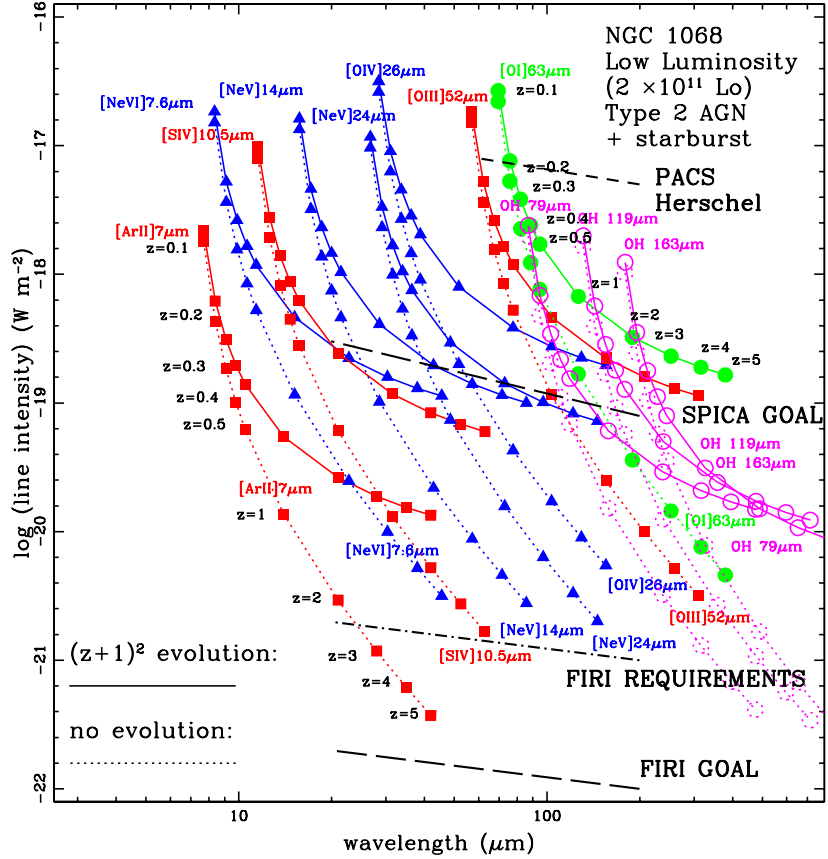}
\caption{Predicted line fluxes as a function of redshift, using as a 
local template the prototypical Seyfert type 2 galaxy NGC1068. 
Squares indicate the fluxes of HII region lines, triangles the fluxes 
of lines emitted by AGN, filled circles the fluxes of the 
[OI]$\mu$m line and open circles the three OH lines 
detected in emission in NGC1068. The solid lines show how 
the line fluxes change with redshift adopting luminosity evolution, 
while dotted lines without any evolution.}
\end{figure}

For simplicity, and to cover as many transitions possible, we use a 
local object which contains both an active nucleus and a starburst, 
the bright prototype Seyfert 2 galaxy NGC1068.
The ISO spectrometers detected in this galaxy many of the lines plotted in Fig. 2 at flux levels
of 5-200 $\times$ 10$^{-16}$ W m$^{-2}$ (Alexander {\em et al.\/} \cite{Ale00}; Spinoglio et al. {\em et al.\/} \cite{spi05}). 
Considering this galaxy as a template object, we computed the line
intensities expected at redshifts ranging from 0.1 to 5. For
simplicity, we adopted an Einstein-De Sitter model Universe, with
$\Omega_{\Lambda}$ = $\Omega_{vac}=0$ and $\Omega_{M}$= 1,
H$_{0}$=75 km s$^{-1}$ Mpc$^{-1}$. The luminosity distances have been derived using:
\begin{equation}
d_{L} (z)= (2c/H_{0})\cdot [1+z - (1+z)^{1/2}]
\end{equation}

The results are reported in Fig. 6, where the line intensities are given in W m$^{-2}$, and the expected sensitivities of spectrometers, such ESI (European SPICA Instrument, Swinyard \cite{swi06}), onboard of the future space observatories, SPICA (Space Infrared Telescope for Cosmology and
Astrophysics, ISAS-JAXA) (Nakagawa \cite{Nag04}; Onaka \& Nakagawa \cite{ON05}) and FIRI (Far-Infrared Interferometer) are also shown.

We have assumed that the line luminosities scale as the bolometric luminosity and we have chosen two cases: \\
A) a luminosity evolution proportional to the (z+1)$^{2}$, consistent with the {\it Spitzer}
results at least up to redshift z=2 (P\'{e}rez-Gonz\'{a}lez {\em et al.\/} \cite{Per05}) ;\\
B) no luminosity evolution.\\

Because the star formation process in galaxies was much more enhanced at z=1-2 than today, we consider reasonable to adopt the model with strong evolution at least for the stellar/HII region lines and, to be conservative, the "no
evolution" one  for the AGN lines.\\
We note that the dependence on different cosmological models is not very strong.
The popular model with $\Omega_{M}$= 0.27, $\Omega_{vac}$=0.73,
H$_{0}$=71 km s$^{-1}$ Mpc$^{-1}$ shows greater dilutions,
increasing with z, by factors of 1.5 for z=0.5 to 2.5 for z=5. In
this case the line intensities of Fig. 6 would decrease by these factors.
 
We conclude that a relatively low luminosity object like NGC1068, with an infrared luminosity of 
2 $\times$ 10$^{11}$ L$_{\odot}$, 
will be detected up to a redshift of z=5 by the cooled 3.5m mirror of the SPICA satellite in a few bright and important diagnostic lines, such as the [OI]63$\mu$m, the   [OIII]52$\mu$m,  [OIV]26$\mu$m, assuming luminosity evolution in the lines.

The fainter AGN lines (like [NeVI]7.6$\mu$m) and the molecular lines of OH, will be detected by SPICA in such an object at z $\sim$ 0.5. For detecting the fainter lines up to z $\sim$ 5, we will have to wait for larger collecting area space telescopes, as the FIRI project, foreseen beyond the next decade.


\end{document}